\documentclass[pra,showpacs,showkeys,twocolumn]{revtex4}
\usepackage{graphicx,amsmath,amsfonts,amssymb}
\usepackage{times}
\begin{document}
\title{Nonmaximally entangled states can be better for quantum correlation distribution and storage}
\author{Xin-Wen Wang,$^{1,2,}$\footnote{xwwang@mail.bnu.edu.cn}
   Shi-Qing Tang,$^1$ Ji-Bing Yuan,$^1$ and Le-Man Kuang$^{1,}$\footnote{lmkuang@hunnu.edu.cn}}
 \affiliation{$^1$ Department of Physics and Key Laboratory of Low-Dimensional Quantum Structures and Quantum Control of Ministry of Education,
  Hunan Normal University, Changsha 410081, People's Republic of China\\
 $^2$Department of Physics and Electronic Information Science,
  Hengyang Normal University, Hengyang 421002, People's Republic of China
 }

\begin{abstract}
For carrying out many quantum information protocols entanglement
must be established in advance between two distant parties.
Practically, inevitable interaction of entangled subsystems with
their environments during distribution and storage will result in
degradation of entanglement. Here we investigate the decoherence of
two-qubit entangled states in the local amplitude noise. We show
that there exists a set of partially entangled states that are more
robust than maximally entangled states in terms of the residual
quantum correlation measured by concurrence, fully entangled
fraction, and quantum discord. This result indicates that
nonmaximally entangled states can outperform maximally entangled
states for quantum correlation distribution and storage under the
amplitude damping. It also educes a notable consequence that the
ordering of states under quantum correlation monotones can be
reversed even by local trace-preserving and completely positive
maps.

\end{abstract}

\pacs{03.67.Hk, 03.65.Yz, 03.67.Pp}

\keywords{Quantum correlation, decoherence, concurrence, fully
entangled fraction, quantum discord}

\maketitle

Establishing quantum entanglement between two distant parties is a
prerequisite for many fundamental tests of quantum theory and
numerous protocols in long-range quantum information processing
\cite{81RMP865}. There are three general ways for providing
long-distance entanglement. One is to prepare a bipartite entangled
pure state in a server and then physically send two subsystems to
two distant parties, respectively. Note that the entanglement of
distributed flying-systems could be transferred to two static
systems \cite{92PRL013602}. The second relies on the successive
interactions of a carrier quantum system with two distant particles
\cite{78PRL3221}. The third is based on the interference of two
photons emitted, respectively, from two distant massive qubits
\cite{83RMP33}. All these methods are usually utilized to create
entanglement between two quantum repeaters or two chosen nodes of a
quantum network \cite{83RMP33,453N1023}.

The above scenarios involve entanglement distribution and storage.
During these processes the entangled systems unavoidably interact
with their environments, respectively, which results in degradation
of established entanglement between the two parties. Investigating
the robustness of entangled states in noise environments may help to
gain some insight into the properties of decoherence and
entanglement, which will provide useful hints for maintaining
entanglement. Many efforts have been devoted to the study of
dynamics of entanglement or quantum correlation for a given initial
two-qubit entangled state in various environments (see, e.g.,
\cite{323S598}). In this Brief Report, we are going to focus on the
question of maximal residual entanglement or quantum correlation of
two-qubit cat-like states for given local noises and decoherence
strengths. It has been shown that the residual entanglement is
proportional to the initial entanglement of the cat-like state in
the case of one-side noise \cite{4NP99}, which implies that the
maximal residual entanglement is attained iff the initial state is
maximally entangled. The purpose of this Brief Report is to
explicitly demonstrate the counterintuitive phenomenon in the case
of two-side noises.

The noise environment considered in this work is the
amplitude-damping environment \cite{Nielsen}. Amplitude-damping
decoherence is applicable to many practical qubit systems, including
vacuum-single-photon qubit with photon loss, atomic qubit with
spontaneous decay, and superconducting qubit with zero-temperature
energy relaxation. The action of an amplitude-damping environment
$\mathcal{E}$ on a single qubit state $\varrho$ can be described as
\begin{equation}
 \varrho\rightarrow\mathcal{E}(\varrho)=M_0\varrho M_0^++M_1\varrho M_1^+,
\end{equation}
where $M_0$ and $M_1$ are the Krauss operators defined by
\begin{eqnarray}
  M_0=\left(
  \begin{array}{cc}
     1 & 0\\
     0 & \sqrt{\overline{d}}
  \end{array}
  \right),~~~~
  M_1=\left(
  \begin{array}{cc}
     0 & \sqrt{d}\\
     0 & 0
  \end{array}
  \right)
\end{eqnarray}
with $0\leqslant d\leqslant1$ and $\overline{d}=1-d$. The
amplitude-damping environment is trace preserving, that is,
$\sum_{j=0,1}M_j^+M_j=I$. Note that $d=0$ denotes the noise-free
case. For $d=1$, the state $\mathcal{E}(\varrho)$ reduces to a
classical state. Therefore, throughout this Brief report, we only
consider the decoherence strength $0<d<1$.

We use three different measures, that is, concurrence
\cite{80PRL2245}, fully entangled fraction (FEF) \cite{54PRA3824},
and quantum discord (QD) \cite{88PRL017901}, to quantify the quantum
correlation of the decoherent state evolving from the initial
entangled state under the action of amplitude-damping environment.
Concurrence is related straightforwardly to the entanglement of
formation and has been considered as a dependable measure of
entanglement for two-qubit entangled states \cite{80PRL2245}. FEF,
which expresses the purity of a mixed state $\rho$, plays a central
role in quantum teleportation and entanglement distillation
\cite{54PRA3824,76PRL722}, and may show different properties from
concurrence \cite{86PRA020304}. QD has been regarded as a more
general measure of nonclassical correlations than entanglement and
even survives entanglement \cite{88PRL017901}, and has been proposed
as the key resource present in certain quantum computational models
and quantum communication tasks without containing much entanglement
\cite{discordapplications}. We show that each of the three measures
reaches its maximum value when the initial two-qubit pure state is a
nonmaximally entangled state but not maximally entangled state. More
importantly, there exists a set of partially entangled states, and
when the initial state belongs to this set, all the aforementioned
quantities of the decoherent state are larger than that in the case
where the initial state is maximally entangled. That is to say, some
partially entangled states can outperform maximally entangled states
for acting as the initial state.

Supposing that two qubits are initially in an entangled state
\begin{equation}
  |\psi\rangle=\sqrt{u}|0\rangle|0\rangle+\sqrt{\overline{u}}e^{i\varphi}|1\rangle|1\rangle,
\end{equation}
where $0\leqslant u\leqslant 1$, $\overline{u}=1-u$, and
$\{|0\rangle,|1\rangle\}$ denotes the computational basis. The
entanglement of $|\psi\rangle$ can be quantified by concurrence. For
a two-qubit state $\rho$ the concurrence is given by
$C(\rho)=\max\{0,\sqrt{\omega_1}-\sqrt{\omega_2}-\sqrt{\omega_3}-\sqrt{\omega_4}\}$,
where $\{\omega_i\}$ are the eigenvalues of
$\rho(\sigma_y\otimes\sigma_y\rho^*\sigma_y\otimes\sigma_y)$ in
decreasing order with $\rho^*$ the complex conjugation of $\rho$ and
$\sigma_y$ the conventional Pauli matrix \cite{80PRL2245}. After a
simple calculation one can obtain
$\mathcal{C}(|\psi\rangle)=2\sqrt{\overline{u}u}$. Obviously,
$\mathcal{C}(|\psi\rangle)$ is equal to unit if $u=1/2$, which means
that $|\psi\rangle$ is a maximally entangled state, denoted by
$|\psi_m\rangle$, in this case. We now consider that each qubit
undergoes an amplitude-damping decoherence during transmitting or
storing. The maximally entangled state $|\psi_m\rangle$ was usually
utilized to implement many quantum information protocols, e.g.,
quantum teleportation \cite{488N185}. The following discussions
suggest that $|\psi\rangle$ with $u\neq 1/2$ could be better than
$|\psi_m\rangle$ for performing these tasks in the amplitude-damping
environment. For convenience in analysis, we assume the decoherence
strengths of two qubits are the same (which makes no difference to
the final conclusion). Then the state $|\psi\rangle$ is degraded
into
\begin{eqnarray}
\label{rho}
 \rho^{d,u}&=&\sum\limits_{j,k=0,1}M_j\otimes
M_k|\psi\rangle\langle\psi|
M_j^+\otimes M_k^+\nonumber\\
&=&\left(
\begin{array}{cccc}
   u+\overline{u}d^2  &  0 & 0 & \overline{d}\sqrt{u\overline{u}}e^{-i\varphi}\\
   0  & \overline{u}\overline{d}d & 0 & 0\\
   0  & 0 & \overline{u}\overline{d}d & 0 \\
 \overline{d} \sqrt{u\overline{u}}e^{i\varphi}& 0 & 0 & \overline{u}\overline{d}^2
\end{array}
\right),
\end{eqnarray}
where $d$ denotes the decoherence strength of each qubit in the
noise environment. The superscripts of $\rho$ denote that the state
$\rho$ is dependent on the parameters $d$ and $u$.

 The concurrence
of $\rho^{d,u}$ is given by
\begin{equation}
\mathcal{C}(\rho^{d,u})=\max\left\{0,2\overline{d}\left(\sqrt{\overline{u}u}-\overline{u}d\right)\right\}.
\end{equation}
It can be seen that $\mathcal{C}(\rho^{d,u})=0$ if
$\sqrt{u/\overline{u}}<d$, which means the occurrence of
entanglement sudden death \cite{97PRL140403}. For a given $d$, the
maximum value of $\mathcal{C}(\rho^{d,u})$ is
\begin{equation}
\label{Cm}
 \mathcal{C}(\rho^{d,u=u_m})=\overline{d}\left(\sqrt{1+d^2}-d\right)
\end{equation}
with
\begin{equation}
\label{um} u_m=\frac{1}{2}+\frac{d}{2\sqrt{1+d^2}}.
\end{equation}
$\mathcal{C}(\rho^{d,u=1/2})<\mathcal{C}(\rho^{d,u=u_m})$ implies an
interesting phenomenon that the decoherent state $\rho^{d,u}$ has
the maximal amount of entanglement if the initial state
$|\psi\rangle$ is an appropriate partially entangled state but not
maximally entangled state. This result is different from that in the
case of one-side noise where the decoherent state has the maximal
amount of entanglement in terms of concurrence when the initial
state is maximally entangled (i.e, $u=1/2$)
\cite{4NP99,86PRA020304}. Particularly, if
\begin{equation}
\label{u'} \frac{1}{2}<u=u'<\frac{1}{2}+\frac{d}{1+d^2},
\end{equation}
the corresponding concurrence is larger than that in the case of
$u=1/2$, that is,
$\mathcal{C}(\rho^{d,u'})>\mathcal{C}(\rho^{d,1/2})$. The above
result implies that the amount of entanglement of $\rho^{d,u}$ is
not monotonic with respect to that of $|\psi\rangle$. For showing
this more clearly, we rewrite $\mathcal{C}(\rho^{d,u})$ as a
function of $\mathcal{C}(|\psi\rangle)$,
\begin{eqnarray}
  \mathcal{C}(\rho^{d,u})&=&\overline{d}\left[C(|\psi\rangle)-d\left(1-\sqrt{1-C(|\psi\rangle)^2}\right)\right]\nonumber\\
      & &\mathrm{for} ~~\frac{1}{2}\leqslant u\leqslant 1,\\
 \mathcal{C}(\rho^{d,u})&=&\max\left\{0,\overline{d}\left[C(|\psi\rangle)-d\left(1+\sqrt{1-C(|\psi\rangle)^2}\right)\right]\right\}\nonumber\\
     & & \mathrm{for} ~~0\leqslant u<\frac{1}{2}.
\end{eqnarray}
Figure 1 shows that when $1/2\leqslant u\leqslant 1$,
$\mathcal{C}(\rho^{d,u})$ is indeed non-monotonic with respect to
$\mathcal{C}(|\psi\rangle)$ for an arbitrarily given $d$. In fact,
$\mathcal{C}(\rho^{d,u})$ takes the extremum
$\mathcal{C}(\rho^{d,u_m})$ on condition that
$\mathcal{C}(|\psi\rangle)$ is equal to $1/\sqrt{1+d^2}$ but not
one.

\begin{figure}
\center
\includegraphics[height=7cm,width=9cm]{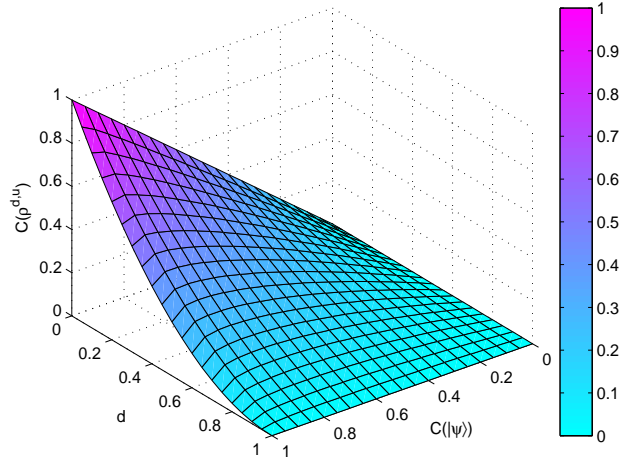}
\caption{(Color online) The residual entanglement
$\mathcal{C}(\rho^{d,u})$ versus the initial entanglement
$\mathcal{C}(|\psi\rangle)$ for $1/2\leqslant u\leqslant 1$.}
\end{figure}

FEF, another important entanglement measure, may behave differently
from concurrence. For example, if the concurrence of a mixed state
$\rho_1$ is larger than the concurrence of a mixed state $\rho_2$,
one cannot conclude that the FEF of $\rho_1$ is also larger than the
FEF of $\rho_2$ \cite{86PRA020304}. FEF is defined as the maximum
overlap of the state $\rho$ with a maximally entangled state,
\begin{equation}
 \mathcal{F}(\rho)=\max\limits_{|\phi\rangle}\langle\phi|\rho|\phi\rangle
\end{equation}
with the maximization taken over all maximally entangled states
$|\phi\rangle$. If and only if $\mathcal{F}(\rho)$ is larger than
1/2, quantum teleportation can exhibit its superiority over the
classical strategy and entanglement distillation can be carried out
effectively in the protocols based on the resource state $\rho$
\cite{54PRA3824,76PRL722}. As a matter of fact, the larger the FEF
$\mathcal{F}(\rho)$ is, the higher teleportation fidelity and
entanglement distillation efficiency can be achieved
\cite{54PRA3824,76PRL722}. For two-qubit systems $\mathcal{F}(\rho)$
can be analytically expressed by \cite{62PRA012311}
\begin{equation}
\mathcal{F}(\rho)=\frac{1+\nu_1+\nu_2-\mathrm{sgn}[\det(\tilde{T})]\nu_3}{4},
\end{equation}
where $\{\nu_i\}$ are the decreasingly ordered singular values of
the $3\times 3$ real matrix
$\tilde{T}=\left[\mathrm{Tr}(\rho\sigma_i\otimes\sigma_j)\right]_{3\times3}$
with $\{\sigma_i,i=1,2,3\}$ the Pauli matrices and
$\mathrm{sgn}[\det(\tilde{T})]$ is the sign of the determinant of
$\tilde{T}$. In our case, the FEF can be calculated to be
\begin{equation}
\mathcal{F}(\rho^{d,u})=\frac{1}{2}+\overline{d}\left(\sqrt{\overline{u}u}-\overline{u}d\right).
\end{equation}
It can be easily verified that $\mathcal{F}(\rho^{d,u})$ gives the
maximum
\begin{equation}
\mathcal{F}(\rho^{d,u=u_m})=\frac{1}{2}+\frac{\overline{d}}{2}\left(\sqrt{1+d^2}-d\right)
\end{equation}
and that $\mathcal{F}(\rho^{d,u'})>\mathcal{F}(\rho^{d,1/2})$.
However, FEF is not an entanglement monotone as it may increase
under trace-preserving local operations and classical communication
(TPLOCC) for mixed states
\cite{62PRA012311,90PRL097901,65PRA022302}. This leads to the advent
of the concept of the maximum achievable FEF given by
\begin{equation}
\mathcal{F}^*(\rho)=\max\limits_{\mathrm{TPLOCC}}\mathcal{F}(\rho)\geqslant
\mathcal{F}(\rho)
\end{equation}
for any $2\otimes2$ density matrix $\rho$ \cite{90PRL097901}.
$\mathcal{F}^*(\rho)$ was shown to be an entanglement monotone
\cite{90PRL097901}. Thus
$\mathcal{F}(\rho^{d,1/2})<\mathcal{F}(\rho^{d,u'})$ does not
necessarily lead to
$\mathcal{F}^*(\rho^{d,1/2})<\mathcal{F}^*(\rho^{d,u'})$. For a
mixed state of two qubits $\rho$, it has been proved that
\cite{66PRA022307}
\begin{equation}
 \mathcal{F}^*(\rho)\leqslant \frac{1+\mathcal{N}(\rho)}{2},
\end{equation}
where $N(\rho)$ denotes the negativity \cite{negativity} of $\rho$
given by $\mathcal{N}(\rho)=\max\{0,-2\lambda_m\}$ with $\lambda_m$
the minimal eigenvalue of the partial transpose of $\rho$ denoted as
$\rho^\Gamma$. The equality is achieved iff the eigenvector
corresponding to the negative eigenvalue of $\rho^\Gamma$ is
maximally entangled \cite{66PRA022307}, and this condition is
equivalent to the condition $\mathcal{N}(\rho)=\mathcal{C}(\rho)$.
In our case,
\begin{equation}
  \mathcal{F}^*(\rho^{d,u})=\mathcal{F}(\rho^{d,u})=\frac{1+\mathcal{N}(\rho^{d,u})}{2}
\end{equation}
if $\sqrt{u/\overline{u}}\geqslant d$ corresponding to
$\mathcal{F}(\rho^{d,u})\geqslant 1/2$. Note that for the values of
$u$ given in Eq.~(\ref{u'}) the condition
$\sqrt{u/\overline{u}}\geqslant d$ is naturally satisfied. In these
cases, therefore, there is no benefit from local processing of the
state $\rho^{d,u}$. In other words, TPLOCC is not required for
obtaining the maximum achievable FEF. Now, we can safely conclude
that $\mathcal{F}^*(\rho^{d,u'})>\mathcal{F}^*(\rho^{d,1/2})$ and
$\mathcal{F}^*(\rho^{d,u})$ gives the extremum
$\mathcal{F}^*(\rho^{d,u_m})=\mathcal{F}(\rho^{d,u_m})$ when the
initial state $|\psi\rangle$ is a partially entangled state with
$u=u_m$. In fact, $\mathcal{F}^*(\rho^{d,u})$ has the same behavior
as $\mathcal{C}(\rho^{d,u})$ for the case
$\sqrt{u/\overline{u}}\geqslant d$.

\begin{figure}
\includegraphics[height=7cm,width=9cm]{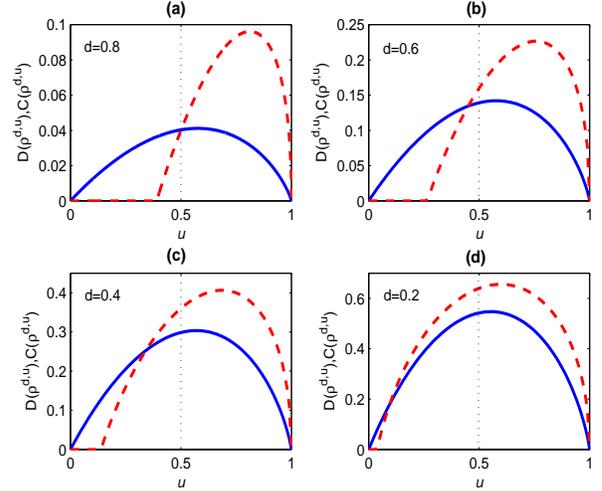}
\caption{(Color online) The concurrence $\mathcal{C}(\rho^{AB})$
(dashed and red line) and the QD $\mathcal{D}(\rho^{AB})$ (solid and
blue line) versus the parameter $u$ of the initial state for a given
decoherence strength $d$. \textbf{(a)} $d=0.8$; \textbf{(b)}
$d=0.6$; \textbf{(c)} $d=0.4$; \textbf{(d)} $d=0.2$.}
\end{figure}

QD is a more general measure of nonclassical correlation than
entanglement \cite{88PRL017901} and may exhibit different features
from concurrence and FEF. Thus, it is necessary to investigate the
QD of the decoherent state $\rho^{d,u}$ in Eq.~(\ref{rho}). For a
given quantum state $\rho^{AB}$ of a composite bipartite system
$AB$, QD is defined as the difference of the quantum mutual
information and the classical correlation \cite{88PRL017901}. The
quantum mutual information is given by
$\mathcal{I}(\rho^{AB})=S(\rho^A)+S(\rho^B)-S(\rho^{AB})$, where
$S(\rho)=-\mathrm{Tr}(\rho\log_2\rho)$ denotes the von Neumann
entropy and $\rho^A$ ($\rho^B$) is the reduced density matrix for
subsystem $A$ ($B$). The classical correlation is quantified by
$\mathcal{J}_A(\rho^{AB})=S(\rho^B)-\min_{E^A_k}\sum_k
p_kS(\rho^{B|k})$, where the minimum is taken over all possible
positive operator valued measures (POVMs) or von Neumann
measurements $\{E_k^A\}$ on subsystem $A$ with
$p_k=\mathrm{Tr}(E_k^A\rho^{AB})$ and
$\rho^{B|k}=\mathrm{Tr}_A(E_k^A\rho^{AB})/p_k$. Then QD is given by
$\mathcal{D}_A(\varrho^{AB})=\mathcal{I}(\rho^{AB})-\mathcal{J}_A(\rho^{AB})$.
Generally, it is very difficult to analytically compute the
classical correlation $\mathcal{J}_A(\rho^{AB})$ due to the
difficulty of finding the optimal measurement for minimizing $\sum_k
p_kS(\rho^{B|k})$. Fortunately, the state $\rho^{d,u}$ in
Eq.~(\ref{rho}) is a particular case of X-states whose QDs have been
intensely investigated \cite{77PRA042303,84PRA042313}. For a real
X-state
\begin{eqnarray}
X^{AB}=\left(
\begin{array}{cccc}
\rho_{00} & 0 & 0 & \rho_{03}\\
0 & \rho_{11} & \rho_{12} & 0\\
0 & \rho_{12} & \rho_{22} & 0\\
\rho_{03} & 0 & 0 & \rho_{33}
\end{array}
\right)
\end{eqnarray}
with $|\rho_{12}+\rho_{03}|\geqslant|\rho_{12}-\rho_{03}|$, it has
been proved \cite{84PRA042313} that the optimal measurement for
$\mathcal{D}_A(X^{AB})$ is $\sigma^A_z$ if
$(|\rho_{12}|+|\rho_{03}|)^2\leqslant(\rho_{00}-\rho_{11})(\rho_{33}-\rho_{22})$
or $\sigma^A_x$ if
$|\sqrt{\rho_{00}\rho_{33}}-\sqrt{\rho_{11}\rho_{22}}|\leqslant|\rho_{12}|+|\rho_{03}|$.
Via a local unitary transformation, which preserves the QD, the
complex factors $e^{\pm i\varphi}$ in the state $\rho^{d,u}$ can be
directly removed and $\rho^{d,u}$ is transformed into a real
X-state. Therefore, we can drop the factors $e^{\pm i\varphi}$ when
we evaluate the QD $\mathcal{D}(\rho^{d,u})$, and the theorem above
is applied to the state $\rho^{d,u}$. Then it can be easily verified
that $\rho^{d,u}$ satisfies the latter condition in the theorem and
the optimal measurement for $\mathcal{D}(\rho^{d,u})$ is $\sigma_x$.
After a straightforward calculation we obtain the QD of such a state
\begin{equation}
\mathcal{D}(\rho^{d,u})=\sum\limits_{i=1}^{4}\lambda_i\log_2\lambda_i-\sum\limits_{j=5}^{8}\lambda_j\log_2\lambda_j,
\end{equation}
where
\begin{eqnarray}
&&\lambda_{1,2}=\overline{u}\overline{d}d,\nonumber\\
&&\lambda_{3,4}=\frac{1}{2}(1-2\overline{u}\overline{d}d)\pm\frac{1}{2}\sqrt{1-4\overline{u}\overline{d}d},\nonumber\\
&&\lambda_{5,6}=\frac{1}{2}\pm\frac{1}{2}\sqrt{1-4\overline{u}\overline{d}d},\nonumber\\
&&\lambda_7=u+\overline{u}d,~~~~\lambda_{8}=\overline{u}\overline{d}.
\end{eqnarray}
Here we have dropped the qubit index in $\mathcal{D}(\rho^{d,u})$
because $\rho^{d,u}$ is invariant under permutation of the two
qubits. Figure 2 shows that for a given $d$ it does not occur in the
case $u=1/2$ that $\mathcal{D}(\rho^{d,u})$ reaches the extremum.
This indicates that if the initial state $|\psi\rangle$ is a
suitable partially entangled state ($u\neq1/2$), the QD
$\mathcal{D}(\rho^{d,u})$ will attain the maximum value.

As shown above, the concurrence, maximum achievable FEF, and QD of
the decoherent state $\rho^{d,u}$ in Eq.~(\ref{rho}) have the same
feature that each of them attains its maximum value iff the initial
state $|\psi\rangle$ is a suitable partially entangled state. That
is, $\mathcal{C}(\rho^{d,u})$, $\mathcal{F}^*(\rho^{d,u})$, and
$\mathcal{D}(\rho^{d,u})$ achieve their extrema iff $u$ is equal to
$u_1=u_m\neq1/2$, $u_2=u_m$, and $u_3\neq1/2$, respectively. It is
worth pointing out that for different values of $d$ different
optimal values of concurrence, FEF, and QD are obtained,
respectively. The corresponding partially entangled states are
different as well. Furthermore, for $u'\in(1/2,u'_0)$ and $u''\in
(1/2,u''_0)$ we have
$\mathcal{C}(\rho^{d,u'})>\mathcal{C}(\rho^{d,1/2})$,
$\mathcal{F}^*(\rho^{d,u'})>\mathcal{F}^*(\rho^{d,1/2})$, and
$\mathcal{D}(\rho^{d,u''})>\mathcal{D}(\rho^{d,1/2})$. Note that
$u_1=u_2\neq u_3$ and $u'_0\neq u''_0$ (see Fig.~2), which provides
a new example of showing the different behaviors of QD and
entanglement. The above results indicate that if the initial state
$|\psi\rangle$ belongs to a suitable set of partially entangled
states, all the concurrence, maximum achievable FEF, and QD of the
decoherent state $\rho^{d,u}$ are larger than that in the case where
$|\psi\rangle$ is maximally entangled. Therefore, we conclude that
partially entangled states can outperform maximally entangled states
for acting as the initial entangled state. It is easy to verify that
all the concurrence, maximum achievable FEF, and QD of the initial
state $|\psi\rangle$ reach their maximum values when $u$=1/2. Thus
the results obtained above also imply that the ordering of states
under concurrence, maximum achievable FEF, and QD can be reversed,
respectively, even though the states undergo identical local
interactions via trace-preserving and completely positive maps.

In conclusion, we have investigated the robustness of two-qubit
cat-like states in the amplitude-damping environment. We
demonstrated that nonmaximally entangled states can be better than
maximally entangled states for quantum correlation distribution and
storage. The obtained results may be of practical importance for
quantum information processing as well as contribute to our
understanding of quantum noises and quantum correlations. By the
way, it has also been reported that nonmaximally entangled states
can outperform maximally entangled states for several quantum tasks
\cite{100PRL110503}.

\begin{acknowledgements}
This work was supported by the NSFC (No.~11004050, No.~11075050, and
No.~11375060), the Key Project of Chinese Ministry of Education
(No.~211119), and the China Postdoctoral Science Foundation funded
project (No.~2013T60769).
\end{acknowledgements}

\end{document}